%% file: main.tex
\documentclass{optica-article}

\journal{opticajournal}

\articletype{Research Article}

\usepackage{hyperref}

\usepackage{braket}
\usepackage{lineno}
\usepackage{siunitx}
\usepackage{subcaption}

\usepackage{fancyhdr}
\begin{document}

\title{Quantifying Effective Heterodyne Detection Efficiency with SI-Traceable Standards}

\author{Luiz Couto Correa Pinto Filho,\authormark{1,*} Jesper B. Christensen,\authormark{1} Anders Brusch,\authormark{1} and Mikael Lassen\authormark{1,*}}

\address{\authormark{1} Danish Fundamental Metrology, Kogle Allé 5, 2970 Hørsholm, Denmark \\ 
}

\email{\authormark{*}lco@dfm.dk and/or ml@dfm.dk}

\begin{abstract*}
Accurate calibration of coherent optical receivers is essential for reliable performance assessment in coherent communications, precision and quantum sensing, and continuous-variable quantum key distribution (CV-QKD), where the effective detection efficiency directly impacts channel parameter estimation. We present a methodology traceable to the International System of Units (SI) to determine the effective heterodyne detection efficiency of balanced receivers using shot-noise-referenced measurements. The protocol relies on two observables acquired with an electrical spectrum analyzer: the heterodyne beat-note power and the local oscillator shot-noise variance, with explicit treatment of the analyzer’s equivalent noise bandwidth (ENBW). The photon flux in the signal path is referenced to SI units via calibrated radiometric standards. We first validate the protocol on a free-space receiver, demonstrating consistency with an independently constructed optical loss chain across a wide range of signal powers and under controlled, calibrated attenuation. Extending the same estimator to a fiber-coupled, polarization-maintaining balanced receiver confirms that the protocol is robust for practical coherent-receiver architectures and intermediate frequencies in the \si{\mega\hertz} range. These results establish a traceable, uncertainty-bounded framework for real-time receiver calibration, providing a practical route for CV-QKD and other coherent optical systems.

\end{abstract*}

\section{Introduction}

Coherent optical receivers underpin a wide range of photonic systems, including coherent communications \cite{tsujino2011quantum,kikuchi2015fundamentals}, optical precision sensing \cite{hansen2001ultrasensitive,bond2016interferometer,akatev2025broadly}, and quantum communication and sensing \cite{madsen2012continuous,fadel2025quantum,bilkis2020real,wang2023accurate,barbieri2022optical,weedbrook2012gaussian}. In continuous-variable quantum key distribution (CV-QKD), receiver calibration is particularly stringent because the effective detection efficiency $\eta$ enters directly into the estimation of channel transmittance and excess noise, and therefore into composable key-rate bounds~\cite{pirandolaAdvancesQuantumCryptography2020,jain2022practical}. A biased estimate of excess noise can lead to overly optimistic performance claims and, in the worst case, operation in insecure regimes \cite{jouguet2013preventing}. Conversely, underestimation of the channel transmittance leads to biased estimates of the achievable communication range and undermines fair benchmarking between systems. This motivates the adoption of calibration procedures based on SI-traceable optical power references, together with comprehensive uncertainty budgets, to enable reproducible and comparable receiver characterization across laboratories \cite{gui2022metrology}.

In CV-QKD, Alice encodes information onto the quadratures of an optical field, which is recovered by Bob using coherent detection. In homodyne detection, a single field quadrature is measured, with the measured quadrature determined by the relative phase between the signal and the local oscillator (LO). In contrast, heterodyne detection enables simultaneous measurement of both quadratures, at the cost of a \SI{3}{\decibel} noise penalty originating from the additional vacuum noise introduced in the measurement process \cite{weedbrook2012gaussian,pirandolaAdvancesQuantumCryptography2020}. In many recent implementations, a local local oscillator (LLO) is generated at the receiver and phase-referenced to the incoming signal using pilot tones or reference pulses \cite{ruiz2024low}, improving security and system robustness compared to transmitting a strong LO through the quantum channel \cite{qi2015generating,jouguet2013experimental}.

Rigorous calibration of homodyne detection efficiency has been addressed in~\cite{zouRigorousCalibrationHomodyne2022}. For balanced heterodyne receivers, however, experimentally validated and SI-traceable protocols are less established \cite{lupo2022quantum}, especially for MHz intermediate frequency (IF) operation where the finite resolution bandwidth (RBW) of an electrical spectrum analyzer (ESA) and its equivalent noise bandwidth (ENBW) must be handled explicitly \cite{NISTWaveformMetrology}. Moreover, the effective heterodyne detection efficiency relevant for receiver benchmarking is not set by a single component parameter: it bundles optical loss and beam-splitter imbalance, mode overlap between signal and LO, photodiode quantum-efficiency \cite{pereira2021impact,NISTDetectorMetrology}, and the contributions of electronic noise and measurement bandwidth.

Here we present and demonstrate an SI-traceable calibration protocol for the effective heterodyne detection efficiency of balanced heterodyne receivers based on shot-noise referenced measurements. The protocol maps measurable quantities, the beat-note power and the LO shot-noise variance measured on an ESA, together with a traceably calibrated signal photon flux, to a single effective efficiency parameter $\eta$ reported with quantified uncertainty. To validate the protocol, we implement a free-space configuration with a separately quantified optical loss-chain estimate and verify agreement within uncertainty. We also introduce calibrated attenuation using neutral-density filters and confirm the expected scaling of $\eta$ with transmission. We finally apply the method to a polarization-maintaining fiber-coupled heterodyne balanced receiver and obtain $\eta = 0.386 \pm 0.012$ ($k=2$) at IFs of \SIlist{20;50;80}{\mega\hertz} for absolute signal powers from the picowatt to the nanowatt range. The results suggest that our protocol provides a practical and traceable framework for heterodyne receiver calibration, with CV-QKD as a representative use case and broader relevance to coherent optical receiver characterization.

\section{Principle of heterodyne efficiency calibration}

\subsection{Heterodyne detection model and definition of effective efficiency}

\begin{figure}[ht]
    \centering
    \includegraphics[width = 0.75\textwidth]{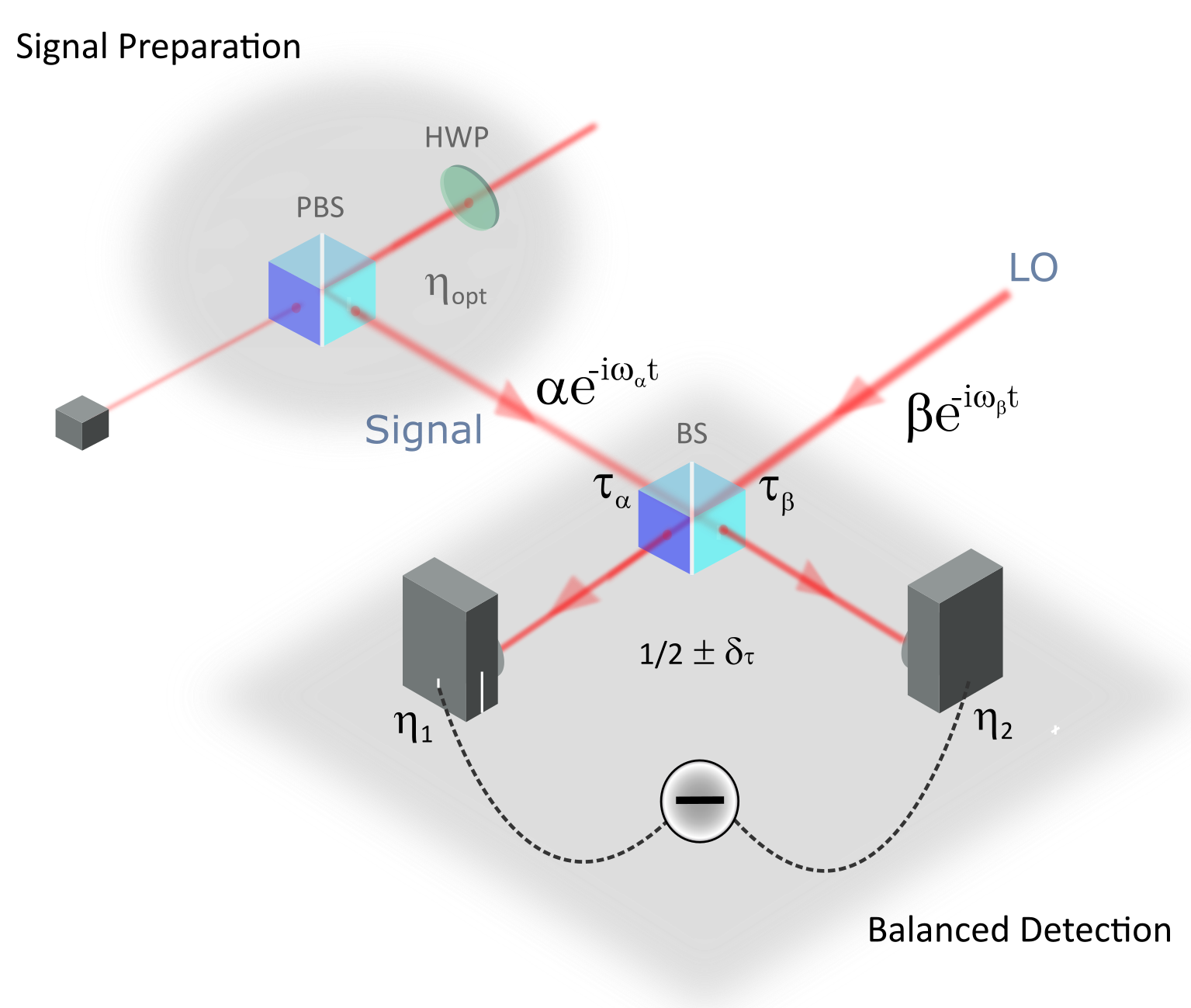}
    \caption{\textbf{Principle of coherent balanced detection and key physical parameters.} The local oscillator (LO), with coherent amplitude $\beta$, and the signal field, with coherent amplitude $\alpha$, are combined at the beam splitter (BS), which may exhibit a small splitting imbalance $(1/2 \pm \delta_{\tau})$ and finite insertion losses described by the coefficients $\tau_{\alpha}$ and $\tau_{\beta}$ for the signal and LO fields, respectively. The heterodyne signal is extracted from the differential photocurrent of two photodetectors with quantum efficiencies $\eta_{1}$ and $\eta_{2}$. Optical losses in the quantum channel are modeled by the effective transmission parameter $\eta_{\mathrm{opt}}$. Losses can be simulated using a half-wave plate (HWP) in combination with a polarizing beam splitter (PBS).}  
    \label{fig:BS}
\end{figure}

We consider the balanced detection configuration shown in Fig.~\ref{fig:BS}. The two input fields incident on the beam splitter (BS) consist of a strong LO with coherent amplitude $\beta$ and angular frequency $\omega_{\beta}$, and a weak signal field with coherent amplitude $\alpha$ $(|\alpha| \ll |\beta|)$ and angular frequency $\omega_{\alpha}$ \cite{bachor2019guide}. The BS is modeled with a nominal 50:50 splitting ratio, including a small imbalance described by $1/2 \pm \delta_{\tau}$, and finite insertion loss represented by non-unit transmission coefficients $\tau_{\alpha}$ and $\tau_{\beta}$ for the signal and LO fields, respectively. Finally, the two detectors are characterized by quantum efficiencies $\eta_1$ and $\eta_2$. Optical losses in the signal arm are modeled by setting the polarizing beam splitter (PBS) to be unbalanced, with transmission $\eta_{\mathrm{opt}}$ \cite{lupo2022quantum,pereira2021impact}. These losses represent quantum channel attenuation and are not directly associated with the intrinsic efficiency of the receiver, but precise knowledge of channel transmission can nevertheless be used to support more comprehensive calibration and parameter estimation of CV‑QKD receivers, ultimately improving the reliability of loss and noise characterization used in secure key rate calculations \cite{jain2022practical}.

In the absence of the signal field, the shot-noise variance, excluding the constant electronic noise contribution, may be found to be:
\begin{equation}
    \left(\Delta I_{-} \right) ^2 = K^2  g^2 F \tau_\beta \vert \beta \vert ^2 (\eta_{av} + \delta_\eta \delta_\tau),
    \label{eq:SNexpression}
\end{equation}
where $K $ is an opto-electronic conversion factor, $g = \langle g \rangle$ is the average electronic gain, $F = \langle g^2 \rangle / \langle g  \rangle ^2 \geq 1$ is the amplifier excess noise,  $\delta_\eta = \eta_1 - \eta_2 $, and $\eta_{av}  = (\eta_1 + \eta_2)/2 $. Since the correction term ($\tau_\beta \vert \beta \vert ^2 \delta_\eta \delta _\tau$) is a second-order term, it will in most cases be small. As an example, consider a 45/55 BS and two detectors with quantum efficiencies \SI{70}{\percent} and \SI{80}{\percent}, then $\delta _\eta \delta _\tau = 0.1 \times 0.05 = 0.005$, which compared to $\eta_{av} = 0.75$ is a relative correction of only \SI{0.6}{\percent} for a rather large imbalance. Thus, in practice $\left( \Delta I_- \right)^2 \approx \tau_\beta \eta_{av} \vert \beta \vert ^2$ is a valid approximation, which only in extreme, and likely impractical, cases results in an error that can exceed \SI{1}{\percent}.  

Now considering the situation with both fields as non-vacuum coherent states, the root-mean-square (RMS) expectation value of the differential photocurrent, excluding DC components, is given as:
\begin{equation}\label{eq:IRMS}
    \langle I_{-} \rangle_{RMS} =  K g \,  \sqrt{2 \left(1 - 4\delta _\tau ^2 \right) \tau_\alpha \tau_\beta \eta _{mm}} \, \,\eta_{av}  \vert \alpha \vert \vert \beta \vert \, ,
\end{equation}
where $\eta_{mm}$ represents the modal field overlap between the signal and the LO. Notably, the beat signal decreases with increasing beam-splitter imbalance. This is then compared with the LO shot-noise variance to provide a signal-to-noise ratio. 
Through the combination of Eqs.~(\ref{eq:SNexpression}) and (\ref{eq:IRMS}), it becomes apparent that we may define the lumped (effective) heterodyne efficiency as
\begin{equation}
    \eta =  (1-4\delta _\tau^2)\,\tau_\alpha\,\eta_{av}\,\eta _{mm}
    = \frac{\langle I_- \rangle _{RMS} ^2}{2  \left( \Delta I_- \right)^2  \vert \alpha \vert ^2}\,,
    \label{eq:etabunched}
\end{equation}
where, for the transimpedance and post-amplifier stages employed in this work, the amplifier noise can be accurately modeled as purely additive, corresponding to a noise factor $F=1$. Eq.~(\ref{eq:etabunched}) indicates that the overall detection efficiency, $\eta$, can be determined through two equivalent procedures: either by independently quantifying the optical losses and imbalance factors on the left-hand side, or by directly computing the ratio on the right-hand side from the measured heterodyne tone and the local oscillator shot-noise variance. In this study, we implement both methods and verify that they yield consistent results within the reported uncertainties.

\subsection{Estimator in terms of experimentally measurable quantities}

We may obtain an estimator for the heterodyne detection efficiency that depends only on experimentally accessible quantities. In our field normalization, the coherent amplitude $\alpha$ corresponds to a photon flux $|\alpha|^{2}$, so if the signal power is $P_{\alpha}=\hbar\omega_{\alpha}\,|\alpha|^{2}$ we can rewrite Eq.~\ref{eq:etabunched} as
\begin{equation}
    \eta =\frac{\hbar \omega_\alpha B_{neq}}{2 P_\alpha} X , \hspace{1 cm} X  = \frac{\langle I_- \rangle ^2_{RMS}}{\left(\Delta  I_- \right) ^2 }.
    \label{eq:etaestimate}
\end{equation}
Here $(\Delta I_{-})^{2}$ is the measured shot-noise variance at the detector output. Because the variance is obtained after a finite-bandwidth measurement chain (e.g., analog filtering, digitization, and any subsequent averaging), it scales linearly with the effective integration bandwidth. We therefore refer to the noise to a \SI{1}{\hertz} bandwidth by using the equivalent noise bandwidth $B_{\mathrm{neq}}$: when $(\Delta I_{-})^{2}$ is evaluated from the measured time series within the experiment's bandwidth, the estimator acquires the compensating factor $B_{\mathrm{neq}}$ shown above so that $\eta$ is independent of the particular filter shape and measurement bandwidth, as must be the case.

Eq.~\ref{eq:etaestimate} thus provides a practical model function for estimating $\eta$ in terms of three measurable quantities: the optical power $P_{\alpha}$, the equivalent noise bandwidth $B_{\mathrm{neq}}$, and the signal-to-noise ratio $X$ extracted from the mean beat note and the shot-noise variance.

\subsection{Uncertainty model}
The standard uncertainty ($k = 1$) of $u({\eta})$, based on Eq.~\ref{eq:etaestimate}, takes the form 
\begin{equation}
    u({\eta}) \approx  {\eta} \sqrt{ \left(\frac{u(P_\alpha)}{P_\alpha}\right)^2 + \left(\frac{u(B_{neq})}{B_{neq}}\right)^2 + \left(\frac{u(X)}{X}\right)^2  } . 
    \label{eq:etauncertainty}
\end{equation}
The mean and standard deviation of $X$ are determined based on repeated measurements (with $P_\alpha$ and $B_{neq}$ kept constant). The equivalent noise bandwidth and its uncertainty are found using narrow frequency tones in the frequency band of interest (\SIrange{10}{100}{\mega\hertz}), after which the mean and relative uncertainty are easily evaluated. With a resolution bandwidth set to \SI{1}{\mega\hertz}, we find an equivalent noise bandwidth of \SI{1.12}{\mega\hertz} (i.e., a filter scale factor of 1.12) with a relative uncertainty of \SI{0.3}{\percent}. The uncertainty of individual optical-power measurements is \SI{0.75}{\percent}, consistent with DFM’s radiometric calibration and measurement capability in the near-infrared (NIR). Traceability is addressed in the Supplementary Material. 

In addition to the statistical  uncertainty in Eq.~\ref{eq:etauncertainty}, the approximation used for the shot-noise variance in Eq.~\ref{eq:SNexpression} leads to a small type B uncertainty. Based on the discussion following Eq.~\ref{eq:SNexpression} and splitting-ratio measurements of the beam splitter used, we choose a worst-case standard uncertainty of $u_B (\eta) = 0.005\eta$, (i.e., a relative uncertainty of \SI{0.5}{\percent}).

\section{Experimental implementation}

\begin{figure}
    \centering
    \includegraphics[width = 0.9\textwidth]{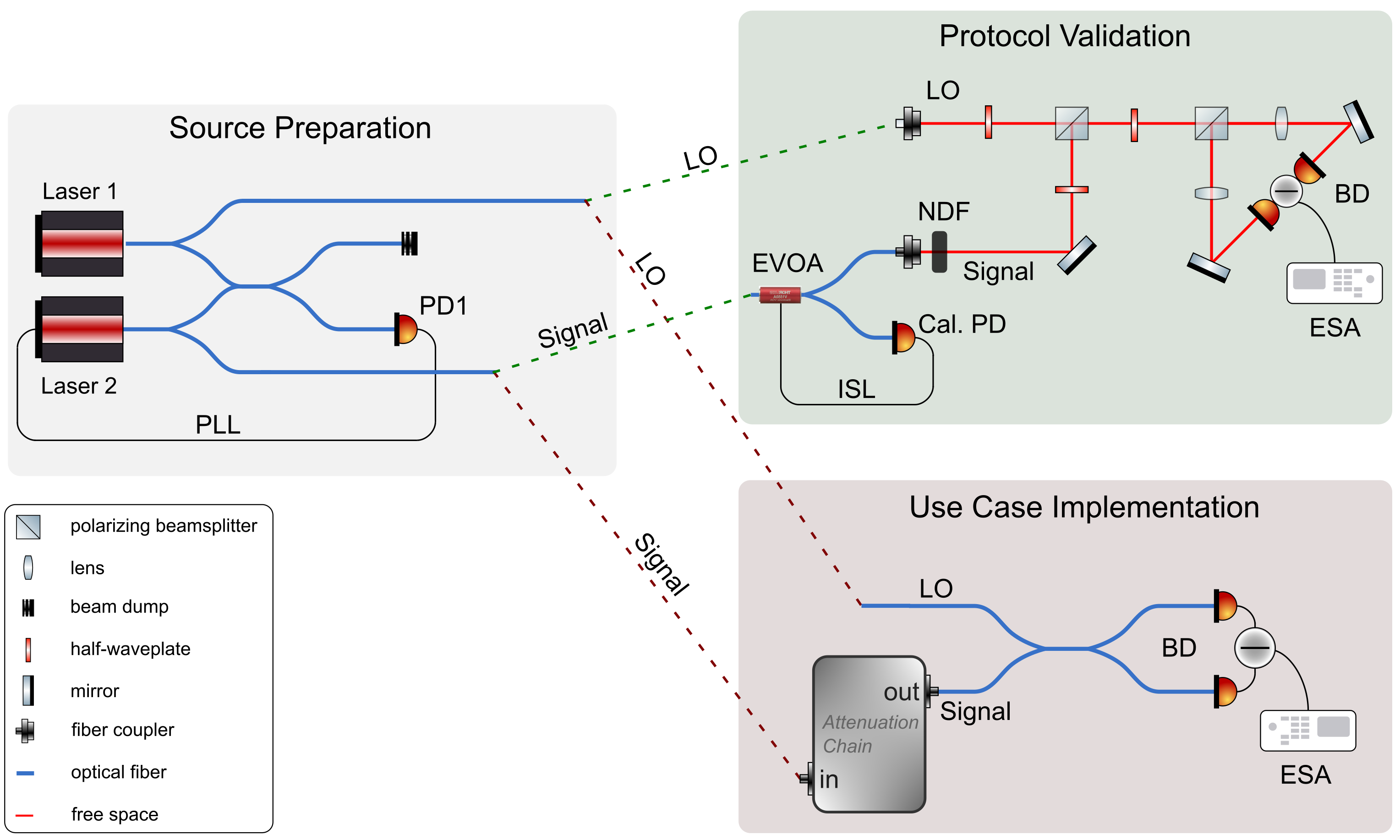}
    \caption{\textbf{Experimental setup.} The experiment is organized in three blocks. \textit{Source preparation:} two narrow-linewidth fiber lasers (primary: DFM’s Stabi$\lambda$aser $1542^{\varepsilon}$ and secondary: NKT Koheras Basik) are frequency-offset locked with a heterodyne optical phase-locked loop (OPLL), using the beat note detected on photodetector PD1. \textit{Protocol validation:} the signal arm is power-stabilized with an electronic variable optical attenuator (EVOA) in an intensity-stabilization loop (ISL), monitored with a calibrated photodetector (Cal.\ PD), and attenuated to nW powers using a calibrated neutral density (ND) filter before being combined with the LO and detected on a balanced detector (BD). \textit{Use-case implementation:} a fully polarization-maintaining fiber implementation where the signal is routed through a calibrated attenuation chain of asymmetric couplers before interfering with the LO in a symmetric coupler; the heterodyne spectrum is measured with a fiber-coupled balanced detector and an ESA.}
    \label{fig:setup_free_space}
\end{figure}

The experiment is organized in three blocks, as summarized in Fig.~\ref{fig:setup_free_space}. We first prepare two phase-coherent optical fields with a controlled frequency offset (Source Preparation). These are then routed either to a free-space setup used to validate the protocol against an independently characterized loss chain (Protocol Validation) or to a fully polarization-maintaining fiber implementation that represents the intended use case (Use Case Implementation). The measurement protocol (electronic-noise, shot-noise, and quadrature/beat-note traces with synchronous power monitoring) is the same in both implementations, and is described separately in the Supplementary Material.

\subsection{Source preparation}

Both implementations share the same optical source preparation. The optical source consists of two narrow-linewidth fiber lasers at \SI{1542}{\nano\meter}. The primary laser (Laser~1) is DFM’s Stabi$\lambda$aser $1542^{\varepsilon}$ \cite{DFMStabilaser1542}, which is acetylene-stabilized and therefore provides both a narrow linewidth and excellent long-term frequency stability. The secondary laser (Laser~2) is an NKT Koheras BASIK. The secondary laser is phase-locked to the primary using a heterodyne optical phase-locked loop (OPLL), with the beat note detected on photodetector PD1. The OPLL is implemented on a RedPitaya STEMlab 125-14 using the \texttt{PyRPL} Python interface, relying on the built-in phase-frequency detector (PFD) and PID controller blocks to generate the error signal and apply feedback to the actuator of the secondary laser. The OPLL offset is set to the intermediate frequency used for the heterodyne measurements: it is fixed to \SI{20}{\mega\hertz} for the free-space protocol-validation dataset, while in the fiber use-case measurements it is set to the IF values used in the receiver characterization (\SIlist{20;50;80}{\mega\hertz}).

In both cases, the primary laser provides the local oscillator (LO) for the balanced coherent receiver, while the secondary laser provides the weaker signal field used for the efficiency calibration.

\subsection{Protocol validation}
The Protocol Validation block in Fig.~\ref{fig:setup_free_space} is a free-space implementation used to validate the protocol against a fully characterized loss chain where all the terms in the left-hand side of Eq.~\ref{eq:etabunched} (optical losses, photodiode quantum efficiency, beam spatial mode overlap, etc.) are independently estimated. For all validation measurements, the OPLL offset is set to \SI{20}{\mega\hertz}, which defines the heterodyne frequency for the acquired spectra.

The signal arm is actively power-stabilized using an electronic variable optical attenuator (EVOA) and a calibrated detector in a feedback loop and then attenuated to the nanowatt level using a calibrated neutral density (ND) filter. The intensity-stabilization loop (ISL) is implemented on the same RedPitaya STEMlab 125-14 used for the OPLL: we use the second RedPitaya input/output channel together with its dedicated PID controller block to generate the ISL error signal from the calibrated monitor detector and drive the EVOA. The LO and signal are combined at a free-space beam splitter and detected with the balanced receiver. The electrical output is sent to a computer-controlled electrical spectrum analyzer (ESA), which records the electronic-noise, shot-noise, and quadrature/beat-note traces (see Supplementary Material). In the free-space measurements, the ESA is operated with fixed center frequency at \SI{20}{\mega\hertz}, a \SI{10}{\mega\hertz} span, and a \SI{1}{\mega\hertz} RBW (with the corresponding ENBW correction characterized in the Supplementary Material), and each stored trace corresponds to an averaged spectrum. Together with the synchronous optical power measurement (from the calibrated detector), this provides the inputs required to evaluate the right-hand side of Eq.~\ref{eq:etabunched} and to compare against the independently determined loss-chain estimate.

\subsection{Use Case Implementation}

The Use Case Implementation block in Fig.~\ref{fig:setup_free_space} is a fully polarization-maintaining (PM) fiber-based realization intended to represent practical coherent-receiver operation. The optical source preparation is identical to Sec.~3.1, but the OPLL offset is set to the IF of interest for each measurement (\SI{20}{\mega\hertz}, \SI{50}{\mega\hertz}, or \SI{80}{\mega\hertz}).

In this configuration, DFM's calibrated attenuation chain of asymmetric couplers is used to traceably attenuate the signal field into the picowatt regime while simultaneously providing optical power stability. The chain implements SI-traceable attenuation using asymmetric PM fiber couplers, yielding a total attenuation of approximately $10^{9}$--$10^{10}$. It also emulates different quantum channel transmissions ($\eta_{\mathrm{opt}}$), enabling controlled testing of the receiver under varying loss conditions. The final signal power is determined from the chain readout, which is calibrated against DFM's radiometric standards to ensure SI traceability.

After transmission through the calibrated attenuation chain, the signal is interfered with the LO on a 50:50 PM coupler and detected using a fiber-coupled balanced receiver module, in which the fibers are permanently fixed. As a result, individual calibration of the photodiodes is not feasible, and the receiver is treated as a single, integrated unit. The ESA acquisition and subsequent signal processing follow the same methodology as employed in the free-space validation measurements, allowing direct application of the estimator to the fiber-coupled receiver.

\section{Results and validation}\label{sec:results}

\subsection{Free-space validation vs independent loss-chain estimate}

The central concept underlying the validation is that the effective heterodyne detection efficiency can be determined via two complementary approaches. In the protocol, the lumped efficiency $\eta_\mathrm{prot}$ is estimated directly from shot-noise-referenced spectral measurements and a traceably calibrated signal power (Eq.~\ref{eq:etaestimate}). Independently, an alternative estimate, $\eta_{\mathrm{sep}}$, can be obtained by separately characterizing the relevant optical and detection losses, including beam-splitter transmissions, photodiode quantum efficiencies, and mode overlap, and combining them into a single effective value. Using the left-hand side of Eq.~\ref{eq:etabunched}, this loss-chain estimate can be expressed as $\eta_{\mathrm{sep}}=(1-4\delta _\tau^2)\,\tau_\alpha\,\eta_{av}\,\eta _{mm}$. 

\subsubsection{Efficiency versus optical power at \SI{20}{\mega\hertz}}

\begin{figure}[ht]
    \centering
    \begin{subfigure}[t]{0.49\textwidth}
        \centering
        \includegraphics[width=\linewidth]{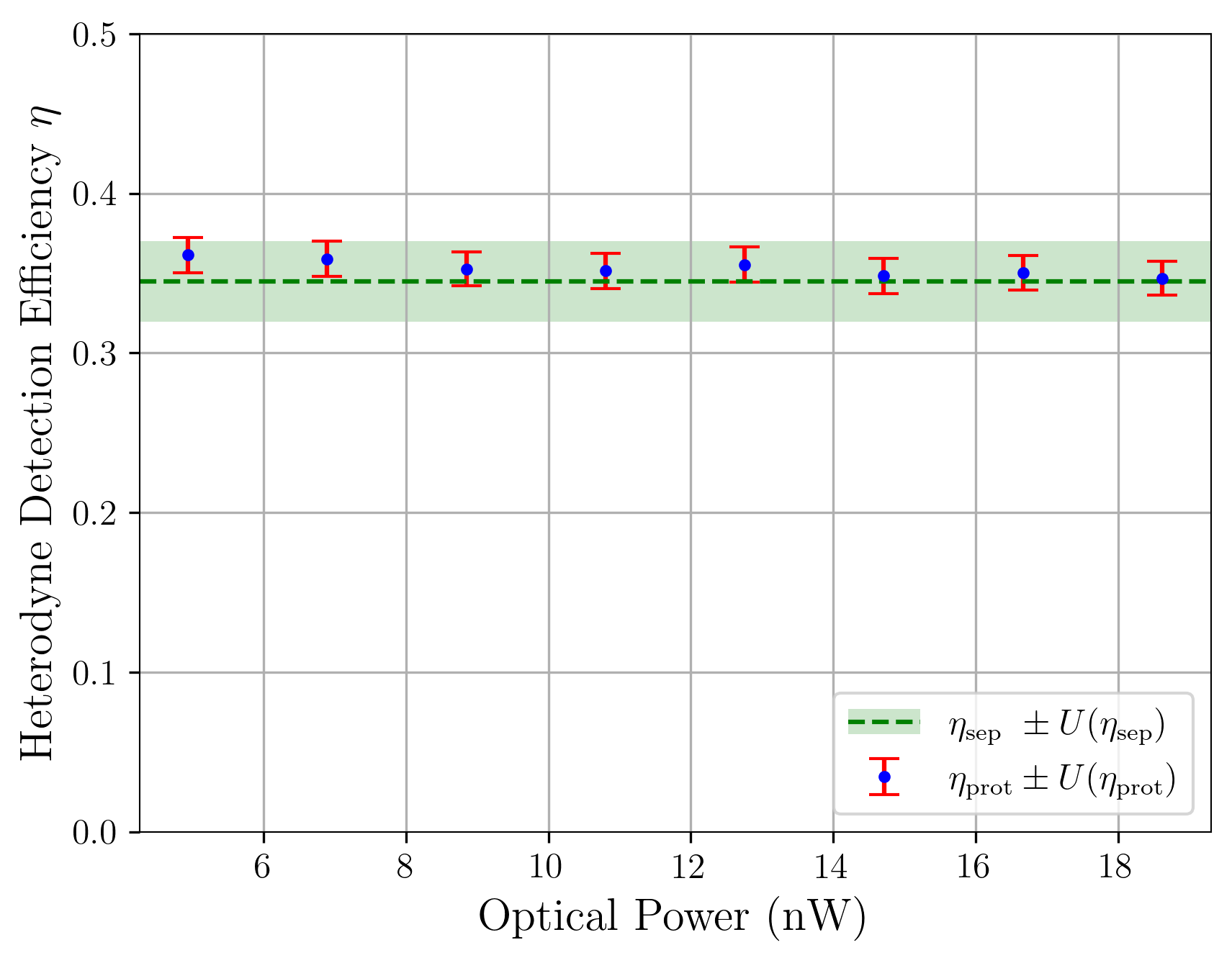}
        \caption{ }
        \label{fig:measurement_free_space}
    \end{subfigure}
    \hfill
    \begin{subfigure}[t]{0.49\textwidth}
        \centering
        \includegraphics[width=\linewidth]{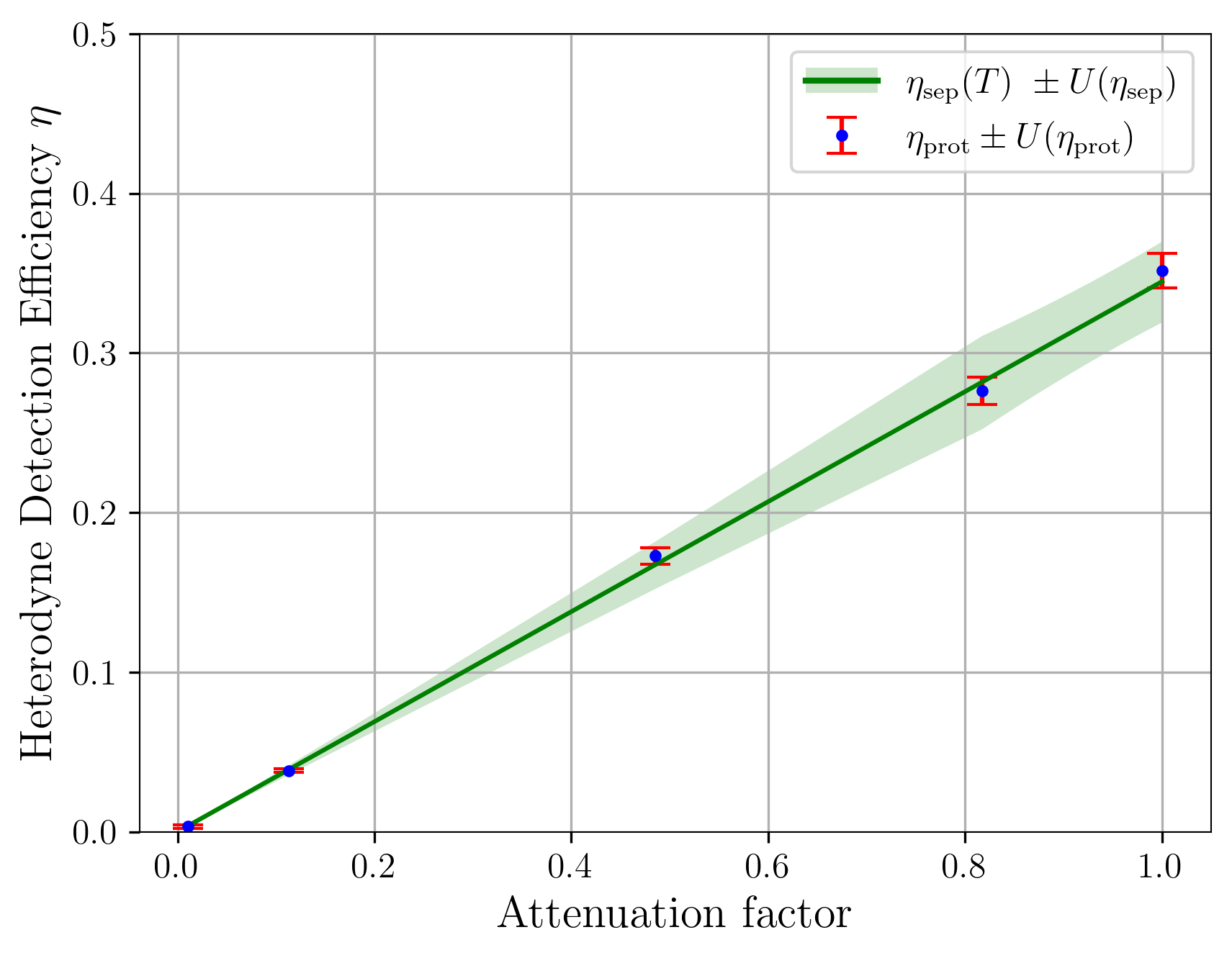}
        \caption{ }
        \label{fig:measurement_free_space_ND}
    \end{subfigure}
    \caption{\textbf{Free-space validation of the protocol.}  Blue dots: protocol-based estimate ($\eta_{\mathrm{prot}}$) with error bars. Green dashed line and band: independent loss-chain estimate ($\eta_{\mathrm{sep}}$). 
    (a) $\eta_{\mathrm{prot}}$ versus signal power with the independent loss-chain estimate $\eta_{\mathrm{sep}}  = 0.345 \pm 0.025$, $k=2$ shown as a band.
    (b) $\eta_{\mathrm{prot}}$ versus calibrated attenuation. The independent prediction is scaled by the measured transmission. In both cases the protocol and independent estimates agree within uncertainties.}
    \label{fig:free_space_validation}
\end{figure}

We first test the protocol-derived efficiency estimate $\eta$ as a function of the absolute signal power $P_\alpha$ in the free-space configuration. For these measurements the IF is fixed to \SI{20}{\mega\hertz} and the LO power is held constant at approximately \SI{1}{m\watt}, while the signal power is varied in the nanowatt range. Fig.~\ref{fig:measurement_free_space} shows the resulting $\eta$ values obtained from repeated acquisitions at each power level. Across a signal power range of 4.9 to 18.6~nW, we observe detection efficiencies between 0.346 and 0.360, with a typical expanded uncertainty of approximately 0.011 ($k=2$). The uncertainty is primarily dominated by contributions from the estimation of $P_\alpha$ and the ENBW correction, rather than by measurement repeatability.

To validate the protocol against an independent reference, we also evaluate an independent efficiency estimate $\eta_{\mathrm{sep}}$ based on a separately quantified loss chain from the calibrated power reference plane to the balanced detector, including the measured transmission of the optical components and the photodiode quantum efficiencies. 
As shown in Fig.~\ref{fig:measurement_free_space}, the protocol-derived efficiencies agree with $\eta_{\mathrm{sep}}=0.345 \pm 0.025$ $(k=2)$ within the stated uncertainties across all the measurements. This agreement indicates that the estimator correctly maps the measured beat-note and shot-noise quantities to an effective heterodyne detection efficiency, and that the radiometric traceability chain for $P_\alpha$ is consistent with the receiver response.

\subsubsection{Efficiency versus attenuation using calibrated ND filters}

As a second, more stringent validation, we deliberately introduce controlled optical losses and verify that the measured efficiency $\eta$ scales as expected. Experimentally, calibrated ND filters are inserted into the signal path, and their transmission $T$ is determined via direct power measurements before and after the filter. For each filter setting, the heterodyne acquisition protocol described above is repeated, and the resulting efficiency values are compared with the independently predicted efficiency $\eta_{\mathrm{sep}}(T)$, obtained by applying the measured transmission to the free-space loss-chain estimate. Fig.~\ref{fig:measurement_free_space_ND} summarizes the results. The measured detection efficiencies decrease with increasing attenuation and remain consistent with the independently calculated model within the stated uncertainties over more than two orders of magnitude in transmission. This validation is complementary to the power sweep shown in Fig.~\ref{fig:measurement_free_space}: whereas the power sweep tests the robustness of the estimator against variations in signal level at a fixed optical configuration, the ND filter measurements confirm that the protocol accurately tracks a known, calibrated change in optical loss.

\subsection{Fiber-coupled receiver demonstration across frequency and power}

We next apply the same estimator to a polarization-maintaining, fiber-coupled balanced heterodyne receiver, representing a practical implementation relevant to modern LLO/pilot architectures for the real-time assessment of receiver efficiency \cite{jain2022practical}. The measurement sequence and data processing remain unchanged, only the optical delivery and receiver configuration differ (see Fig.~\ref{fig:setup_free_space}). Calibrations are performed at intermediate frequencies of \SIlist{20;50;80}{\mega\hertz}, while the absolute signal power is varied across the picowatt to nanowatt range.

\begin{figure}[ht]
    \centering
    \includegraphics[scale=0.6]{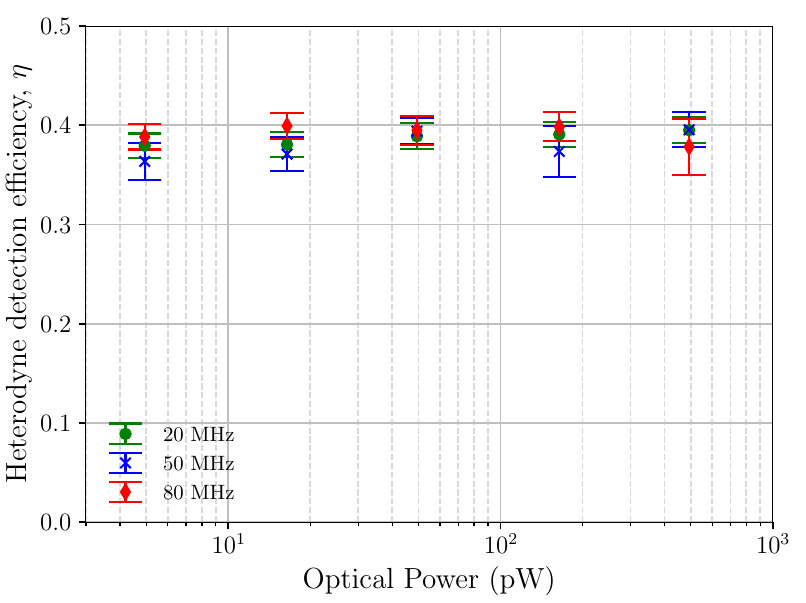}
    \caption{\textbf{Traceable calibration of the lumped heterodyne detection efficiency for the fiber-coupled receiver.} The heterodyne efficiencies derived from the protocol are shown for three intermediate frequencies (IF) (\SIlist{20;50;80}{\mega\hertz}). Within the stated uncertainties, the measured efficiencies are consistent across the investigated IF range and signal powers, yielding a representative value of $\eta = 0.386 \pm 0.012$ ($k=2$).}
    \label{fig:measurement1}
\end{figure}

Fig.~\ref{fig:measurement1} shows the protocol-derived heterodyne efficiencies for the three IF frequencies. Within the stated uncertainties, the measured efficiencies are consistent across the investigated IF range and signal powers, yielding a representative value of $\eta = 0.386 \pm 0.012$ ($k=2$). These results demonstrate that the protocol can be applied directly to fiber-based coherent receivers operating at MHz IF, which is the regime commonly used for pilot-tone-assisted LLO CV-QKD systems and related coherent detection architectures.

\section{Discussion}

The objective of this work was not to optimize a specific receiver design, but rather to provide a practical methodology for assigning a traceable, uncertainty-bounded value to the effective heterodyne detection efficiency of a coherent receiver. The results presented in Sec.~\ref{sec:results} demonstrate that the protocol is robust across a wide range of signal powers, agrees with an independently constructed loss-chain estimate in the free-space implementation, and remains consistent under intentionally applied, calibrated attenuation. Extending the same estimator to a fiber-coupled balanced receiver further confirms that the approach is directly applicable to practical coherent-receiver architectures and to intermediate frequencies in the MHz regime.

A key technical point is the treatment of measurement bandwidth when the shot-noise reference is read out with an electrical spectrum analyzer. Because the RBW filter is not perfectly rectangular, the effective integration bandwidth differs from the nominal RBW and must be accounted for via the equivalent noise bandwidth (ENBW). Neglecting this correction introduces a fixed scale error in the inferred shot-noise level, which maps directly into a bias on $\eta$. Once characterized for a given analyzer configuration, the ENBW correction becomes a fixed contribution to the uncertainty budget and is essential for meaningful comparison across RBW settings, instruments, or laboratories.

CV-QKD provides a stringent and timely use case for the protocol. In parameter estimation, the detection efficiency enters the effective transmittance and therefore affects the inferred channel transmittance and excess noise, with direct impact on composable key-rate bounds. The present protocol offers a concrete route to reporting receiver efficiencies that are SI-traceable and accompanied by quantified uncertainties, rather than being treated as loosely defined effective losses. In addition, the attenuation-validation experiment provides a direct experimental check that the inferred efficiency responds correctly to a controlled, calibrated change in optical loss, closely mirroring how known link losses are handled in system models.

From a practical standpoint, the main limitations are straightforward. The spectrum analyzer must be operated in its linear regime to avoid biasing the extracted beat-note power, and sufficient dynamic range is required to separate electronic noise from optical shot-noise. The extraction of the spectral ratio $X$ also depends on a consistent choice of the frequency window around the heterodyne tone: the window should be wide enough to average the noise, but narrow enough to avoid unrelated spectral features. On the optical side, the protocol assumes that the LO dominates the noise in the measurement band so that the LO-only trace represents shot-noise; excess laser intensity noise or residual modulation in-band would require explicit treatment. Finally, slow drifts in mode overlap or polarization can increase the repeatability contribution if acquisitions are not performed under stable conditions.

Looking ahead, the protocol is well suited to extensions relevant to deployment and standardization. Once the ENBW correction and power-monitor calibration are established, the remaining measurement steps can be automated and used for periodic or in-situ receiver self-checks. More broadly, because the estimator is expressed in terms of SI-traceable optical power and directly measured electrical quantities, the method is naturally suited to inter-laboratory comparisons. Although we focus here on balanced heterodyne receivers, the same shot-noise-referenced approach combined with a traceable photon-flux calibration can be adapted to other coherent-receiver configurations, as long as the measurement bandwidth and instrument response are properly characterized.

\section{Conclusion}

We have presented and experimentally demonstrated a practical, SI-traceable protocol for calibrating the effective heterodyne detection efficiency of balanced heterodyne receivers. The method is based on the estimator in Eq.~(\ref{eq:etaestimate}), which relates a shot-noise-referenced spectral measurement (beat-note power and local-oscillator shot-noise variance) to the calibrated signal photon flux. A key element of the implementation is the correction from the spectrum-analyzer RBW to its equivalent noise bandwidth, which enables a consistent normalization of the measured noise and supports meaningful comparison across settings and instruments.

The protocol was validated in a free-space implementation by comparison to an independently constructed loss-chain estimate. As shown in Fig.~\ref{fig:measurement_free_space}, the protocol-derived efficiencies are stable across all measurements at a \SI{20}{\mega\hertz} IF and agree with the independent estimate within the stated uncertainties. Introducing controlled, calibrated attenuation using ND filters provides a second validation axis: the measured efficiencies follow the expected transmission scaling and remain consistent with the independent model within uncertainty (Fig.~\ref{fig:measurement_free_space_ND}). We further applied the same estimator to a polarization-maintaining fiber-coupled balanced heterodyne receiver and obtained a representative efficiency of $\eta = 0.386 \pm 0.012$ ($k=2$) at IF frequencies of \SIlist{20;50;80}{\mega\hertz} over signal powers from the picowatt to nanowatt range (Fig.~\ref{fig:measurement1}).

Overall, this work provides a traceable and uncertainty-bounded framework for heterodyne-receiver efficiency calibration that is directly applicable to coherent-receiver benchmarking, with CV-QKD as a representative use case. We expect that the combination of a clear estimator, explicit bandwidth correction, and a complete uncertainty budget will facilitate reproducible receiver characterization and support future inter-laboratory comparisons and standardization efforts.

\section*{Funding}
The project is supported by funds from Danish Agency for Higher Education and Science and by the Innovation Fund Denmark under the Grand Solutions program (CryptQ: 0175-00018B and CyberQ: 3200-00035B).

\section*{Acknowledgments}
We thank Tobias Gehring and Nitin Jain for valuable discussions related to the testing and evaluation of our protocol.

\section*{Disclosures}
The authors declare no conflicts of interest.

\section*{Data availability}
Data underlying the results presented in this paper are not publicly available at this time but may be obtained from the authors upon reasonable request.

\bibliography{references.bib}

\clearpage 

\appendix

% Supplement numbering
\setcounter{equation}{0}
\setcounter{figure}{0}
\setcounter{table}{0}

\renewcommand{\theequation}{S\arabic{equation}}
\renewcommand{\thefigure}{S\arabic{figure}}
\renewcommand{\thetable}{S\arabic{table}}

\renewcommand{\theHequation}{S\arabic{equation}}
\renewcommand{\theHfigure}{S\arabic{figure}}
\renewcommand{\theHtable}{S\arabic{table}}

\input{main_supplement}

\end{document}

%% file: main_supplement.tex
\section*{SUPPLEMENTARY MATERIAL}

\section{Theoretical model}

Consider a balanced-detector configuration where the two possible input fields to the beam splitter are a local oscillator (LO) with coherent amplitude $\beta$ at angular frequency $\omega_\beta$, and a weak signal field with coherent amplitude $\alpha \, (\alpha \ll \beta)$ at angular frequency $\omega_\alpha$. The beam splitter is modeled with power transmission and reflection coefficients $T=1/2+\delta_\tau$ and $R=1/2-\delta_\tau$, respectively, and with insertion loss represented by field transmissions $\tau_\alpha$ and $\tau_\beta$ for the signal and LO paths. The photodetectors have quantum efficiencies $\eta_1$ and $\eta_2$. Optical loss in the signal arm can be modeled by an additional imbalanced beam splitter.

We define the usual balanced lossless beam splitter operation as:
\begin{equation}
\begin{pmatrix}
\hat{c}\\
\hat{d}
\end{pmatrix}
=
\frac{1}{\sqrt{2}}
\begin{pmatrix}
1 & 1\\
1 & -1
\end{pmatrix}
\begin{pmatrix}
\hat{a}\\
\hat{b}
\end{pmatrix}
\end{equation}

For an imbalanced beam splitter (BS) with imbalance $\delta_\tau$:
\begin{equation}
\begin{pmatrix}
\hat{c}\\
\hat{d}
\end{pmatrix}
=
\begin{pmatrix}
t & r\\
r & -t
\end{pmatrix}
\begin{pmatrix}
\hat{a}\\
\hat{b}
\end{pmatrix}
\end{equation}
\begin{equation}
r=\sqrt{\frac{1}{2}-\delta_\tau},
\qquad
t=\sqrt{\frac{1}{2}+\delta_\tau}
\end{equation}

\subsection*{Adding losses and efficiencies}

We model input losses as a perfect BS with transmittance $\tau_i$ and vacuum as the input at the
other port:
\begin{equation}
\hat{a}'=\sqrt{\tau_\alpha}\,\hat{a}+\sqrt{1-\tau_\alpha}\,\hat{v}_\alpha
\end{equation}
\begin{equation}
\hat{b}'=\sqrt{\tau_\beta}\,\hat{b}+\sqrt{1-\tau_\beta}\,\hat{v}_\beta
\end{equation}
where $\hat{v}_\alpha$ and $\hat{v}_\beta$ are vacuum operators.
This way we may write:
\begin{equation}
\begin{pmatrix}
\hat{c}_0\\
\hat{d}_0
\end{pmatrix}
=
\begin{pmatrix}
t & r\\
r & -t
\end{pmatrix}
\begin{pmatrix}
\hat{a}'\\
\hat{b}'
\end{pmatrix}.
\end{equation}

We add the detection efficiency in a similar fashion, by adding another vacuum term for each detection channel ($\hat{v}_1$ and $\hat{v}_2$):
\begin{equation}
\hat{c}=\sqrt{\eta_2}\,\hat{c}_0+\sqrt{1-\eta_2}\,\hat{v}_2
\end{equation}
\begin{equation}
\hat{d}=\sqrt{\eta_1}\,\hat{d}_0+\sqrt{1-\eta_1}\,\hat{v}_1
\end{equation}

Substituting and solving for $\hat{c}$ and $\hat{d}$:
\begin{equation}
\hat{c}_0
=
t\left(\sqrt{\tau_\alpha}\,\hat{a}+\sqrt{1-\tau_\alpha}\,\hat{v}_\alpha\right)
+
r\left(\sqrt{\tau_\beta}\,\hat{b}+\sqrt{1-\tau_\beta}\,\hat{v}_\beta\right)
\end{equation}
\begin{equation}
\hat{d}_0
=
r\left(\sqrt{\tau_\alpha}\,\hat{a}+\sqrt{1-\tau_\alpha}\,\hat{v}_\alpha\right)
-
t\left(\sqrt{\tau_\beta}\,\hat{b}+\sqrt{1-\tau_\beta}\,\hat{v}_\beta\right)
\end{equation}

\begin{align}
\hat{c}
&=
\sqrt{\eta_2}\left[t\sqrt{\tau_\alpha}\,\hat{a}+r\sqrt{\tau_\beta}\,\hat{b}\right]
+\hat{v}_c \\
\hat{v}_c
&=
\sqrt{\eta_2}\left[t\sqrt{1-\tau_\alpha}\,\hat{v}_\alpha+r\sqrt{1-\tau_\beta}\,\hat{v}_\beta\right]
+\sqrt{1-\eta_2}\,\hat{v}_2
\end{align}

\begin{align}
\hat{d}
&=
\sqrt{\eta_1}\left[r\sqrt{\tau_\alpha}\,\hat{a}-t\sqrt{\tau_\beta}\,\hat{b}\right]
+\hat{v}_d \\
\hat{v}_d
&=
\sqrt{\eta_1}\left[r\sqrt{1-\tau_\alpha}\,\hat{v}_\alpha-t\sqrt{1-\tau_\beta}\,\hat{v}_\beta\right]
+\sqrt{1-\eta_1}\,\hat{v}_1
\end{align}

We now recover the photon number operators for $\hat{c}$ and $\hat{d}$:
\begin{equation}
\hat{n}_c=\hat{c}^\dagger \hat{c}
=
\eta_2\left[
t^2\tau_\alpha\,\hat{a}^\dagger \hat{a}
+
r^2\tau_\beta\,\hat{b}^\dagger \hat{b}
+
tr\sqrt{\tau_\alpha\tau_\beta}\,(\hat{a}^\dagger \hat{b}+\hat{a}\hat{b}^\dagger)
\right]
+\hat{n}_{c,\text{vac}}
\end{equation}

\begin{equation}
\hat{n}_d=\hat{d}^\dagger \hat{d}
=
\eta_1\left[
r^2\tau_\alpha\,\hat{a}^\dagger \hat{a}
+
t^2\tau_\beta\,\hat{b}^\dagger \hat{b}
-
tr\sqrt{\tau_\alpha\tau_\beta}\,(\hat{a}^\dagger \hat{b}+\hat{a}\hat{b}^\dagger)
\right]
+\hat{n}_{d,\text{vac}}
\end{equation}
where $\hat{n}_{c,\text{vac}}$ and $\hat{n}_{d,\text{vac}}$ represent the combined vacuum terms.

\subsection*{The difference photocurrent operator}
We define the scaled difference in photocurrent as:
\begin{equation}\label{eq:I_def}
\hat{I} = K g \,(\hat{n}_c - \hat{n}_d)
\end{equation}
where $K$ is an opto-electronic conversion factor,
$g=\langle g\rangle$ is the average electronic gain and
\begin{equation}
F=\frac{\langle g^2\rangle}{\langle g\rangle^2}\ge 1
\end{equation}
is the amplifier excess noise. We can then write an expression for the first moment of the photocurrent operator $\langle \hat{I}\rangle$:
\begin{equation}\label{eq:I_first_moment}
\langle \hat{I}\rangle = K g \left(\langle \hat{n}_c\rangle - \langle \hat{n}_d\rangle\right).
\end{equation}

\subsection*{The shot-noise variance}

In order to compute the shot-noise variance, we turn to the fluctuations of the photocurrent operator $\delta\hat{I}$. If we write the photocurrent operator as $\hat{I}=\langle \hat{I} \rangle + \delta\hat{I}$, we can define the fluctuations as:
\begin{equation}
    \delta\hat{I}= \hat{I} - \langle \hat{I} \rangle.
\end{equation}
Then from the definition of the photocurrent operator (Eq. \ref{eq:I_def}) we have:
\begin{align}
    \delta\hat{I} &= K g \,(\hat{n}_c - \hat{n}_d) - K \langle g \rangle \,(\langle\hat{n}_c \rangle- \langle \hat{n}_d\rangle) \nonumber \\
                  &= K g \,\left[ (\hat{n}_c-\langle\hat{n}_c \rangle) - (\hat{n}_d - \langle \hat{n}_d\rangle) \right] \nonumber \\
                  &=  K g \,(\delta\hat{n}_c - \delta\hat{n}_d)
\end{align}
where we define the fluctuations of an arbitrary photon number operator as $\delta \hat{n}_i = \hat{n}_i - \langle \hat{n}_i \rangle$ ($i=c,d$).

To compute the shot-noise variance:
\begin{equation}\label{eq:variance}
\left(\Delta I_-\right)^2  \equiv \langle \delta\hat{I}^2\rangle - \langle \delta\hat{I}\rangle^2
\end{equation}
we need the first and second moments of the fluctuation operator, $\langle \delta\hat{I}\rangle$ and $\langle \delta\hat{I}^2\rangle$. From the definition of the fluctuation operator, the first moment $\langle \delta\hat{I} \rangle = 0$. For the second moment we have that:
\begin{equation}
\langle \delta\hat{I}^2\rangle = K^2 \langle g^2 \rangle \left(\langle \delta\hat{n}_c^{\,2}\rangle + \langle \delta\hat{n}_d^{\,2}\rangle - 2\langle \delta\hat{n}_c \delta\hat{n}_d\rangle\right) = K^2 g^2 F \left(\langle \delta\hat{n}_c^{\,2}\rangle + \langle \delta\hat{n}_d^{\,2}\rangle - 2\langle \delta\hat{n}_c \delta\hat{n}_d\rangle\right).
\end{equation}
We can show that $\langle \delta\hat{n}_c^2\rangle=\langle \hat{n}_c\rangle$, $\langle \delta\hat{n}_d^2\rangle=\langle \hat{n}_d\rangle$ and $\langle \delta\hat{n}_c \delta\hat{n}_d\rangle = 0$.
Then the expression for the shot-noise variance becomes:
\begin{equation}\label{eq:variance_nc_nd}
    \left(\Delta I_-\right)^2 = K^2 g^2 F \left(\langle \hat{n}_c\rangle + \langle \hat{n}_d\rangle \right)
\end{equation}

Because we are interested in the shot-noise variance, the signal is a vacuum state and the LO is a strong
coherent field such that:
\begin{equation}
\hat{a}\lvert 0\rangle = 0
\qquad
\hat{b}\lvert \beta\rangle = \beta \lvert \beta\rangle \, .
\end{equation}
Then,
\begin{equation}\label{eq:nc}
\langle \hat{n}_c\rangle
=
\eta_2\left(\frac{1}{2}-\delta_\tau\right)\tau_\beta\,|\beta|^2
\end{equation}
\begin{equation}\label{eq:nd}
\langle \hat{n}_d\rangle
=
\eta_1\left(\frac{1}{2}+\delta_\tau\right)\tau_\beta\,|\beta|^2 \, .
\end{equation}
Finally, using Eqs.~(\ref{eq:nc}) and (\ref{eq:nd}) in Eq.~(\ref{eq:variance_nc_nd}), we obtain:
\begin{equation}
\left(\Delta I_{-} \right) ^2  = K^2 g^2 F\,\tau_\beta |\beta|^2
\left[
\eta_1\left(\frac{1}{2}+\delta_\tau\right)
+
\eta_2\left(\frac{1}{2}-\delta_\tau\right)
\right]\, .
\end{equation}

\subsection*{Difference photocurrent for two coherent non-vacuum states}

In this case the signal and the LO are strong coherent fields:
\begin{equation}
\hat{a}\lvert \alpha \rangle = \alpha \lvert \alpha \rangle
\qquad
\hat{b}\lvert \beta\rangle = \beta \lvert \beta\rangle
\end{equation}
Then:
\begin{align}
\langle \hat{n}_c\rangle
&=
\eta_2\Big[
t^2 \tau_\alpha |\alpha|^2
+
r^2 \tau_\beta |\beta|^2
+
tr\sqrt{\tau_\alpha\tau_\beta}\,(\alpha^*\beta+\alpha\beta^*)
\Big],
\\[4pt]
\langle \hat{n}_d\rangle
&=
\eta_1\Big[
r^2 \tau_\alpha |\alpha|^2
+
t^2 \tau_\beta |\beta|^2
-
tr\sqrt{\tau_\alpha\tau_\beta}\,(\alpha^*\beta+\alpha\beta^*)
\Big].
\end{align}

\begin{align}
\langle \hat{n}_c\rangle-\langle \hat{n}_d\rangle
&=
\tau_\alpha |\alpha|^2
\Big[
\eta_2\Big(\tfrac{1}{2}+\delta_\tau\Big)
-
\eta_1\Big(\tfrac{1}{2}-\delta_\tau\Big)
\Big]
\nonumber\\
&\quad
+
\tau_\beta |\beta|^2
\Big[
\eta_2\Big(\tfrac{1}{2}-\delta_\tau\Big)
-
\eta_1\Big(\tfrac{1}{2}+\delta_\tau\Big)
\Big]
\nonumber\\
&\quad
+
\sqrt{\left(\tfrac{1}{4}-\delta_\tau^{2}\right)\tau_\alpha\tau_\beta}\,
(\alpha^*\beta+\alpha\beta^*)\,(\eta_1+\eta_2).
\end{align}

\begin{equation}
\delta_\eta \equiv \eta_1-\eta_2,
\qquad
\eta_{\mathrm{av}} \equiv \frac{\eta_1+\eta_2}{2}.
\end{equation}

\begin{equation}
\alpha = |\alpha|e^{-i\omega_\alpha t},
\qquad
\beta = |\beta|e^{-i\omega_\beta t}.
\end{equation}

\begin{align}
\langle \hat{n}_c\rangle-\langle \hat{n}_d\rangle
&=
\tau_\alpha |\alpha|^2\Big[
-\tfrac{1}{2}(\eta_1-\eta_2)+(\eta_1+\eta_2)\delta_\tau
\Big]
\nonumber\\
&\quad
+
\tau_\beta |\beta|^2\Big[
-\tfrac{1}{2}(\eta_1-\eta_2)-(\eta_1+\eta_2)\delta_\tau
\Big]
\nonumber\\
&\quad
+
2\sqrt{\left(\tfrac{1}{4}-\delta_\tau^{2}\right)\tau_\alpha\tau_\beta}\,
\eta_{\mathrm{av}}\,|\alpha||\beta|
\left(
e^{i(\omega_\alpha-\omega_\beta)t}+e^{-i(\omega_\alpha-\omega_\beta)t}
\right)
\\[6pt]
&=
\tau_\alpha |\alpha|^2
\left(
2\eta_{\mathrm{av}}\delta_\tau-\frac{\delta_\eta}{2}
\right)
-
\tau_\beta |\beta|^2
\left(
2\eta_{\mathrm{av}}\delta_\tau+\frac{\delta_\eta}{2}
\right)
\nonumber\\
&\quad
+
4\sqrt{\left(\tfrac{1}{4}-\delta_\tau^{2}\right)\tau_\alpha\tau_\beta}\,
\eta_{\mathrm{av}}\,|\alpha||\beta|\,
\cos\!\big((\omega_\alpha-\omega_\beta)t\big).
\end{align}
If we disregard the DC (non-oscillating) components and substitute into Eq.~(\ref{eq:I_first_moment}), we obtain:
\begin{equation}
    \langle \hat{I} \rangle = 4 K g \, \sqrt{\left(\tfrac{1}{4}-\delta_\tau^{2}\right)\tau_\alpha\tau_\beta}\,
\eta_{\mathrm{av}}\,|\alpha||\beta|\,
\cos\!\big((\omega_\alpha-\omega_\beta)t\big).
\end{equation}

\subsection*{Effect of spatial mode overlap}

Imperfect spatial overlap between the signal and LO reduces their interference. We introduce the overlap
\begin{equation}
\gamma = \int d^2\mathbf r\, u_\alpha^*(\mathbf r)\,u_\beta(\mathbf r),
\qquad 0 \le |\gamma| \le 1,
\end{equation}
so only the overlapping mode contributes to the heterodyne term. $u_\alpha(\mathbf r)$ and $u_\beta(\mathbf r)$ are normalized spatial mode functions. The beat amplitude scales by $|\gamma|$ (equivalently, beat power by $\eta_{mm}=|\gamma|^2$), giving
\begin{equation}
\langle \hat{I} \rangle
=
4 K g \, 
\sqrt{\left(\tfrac{1}{4}-\delta_\tau^{2}\right)\tau_\alpha\tau_\beta\eta_{mm}}\,
\eta_{\mathrm{av}}\,|\alpha||\beta|\,
\cos\!\big((\omega_\alpha-\omega_\beta)t\big).
\end{equation}
Thus, mode mismatch can be folded into the overall efficiency as a multiplicative factor $\eta_{mm}$.

\section{Measurement protocol and acquisition sequence}

For each measurement setting (given IF frequency, analyzer settings, and signal-power level), the acquisition is organized as a short, repeatable sequence that separates the relevant noise references from the beat-note measurement, while keeping the LO conditions unchanged. The protocol is designed so that each saved dataset contains all quantities needed to evaluate
\begin{equation}
    \eta =\frac{\hbar \omega_\alpha B_{neq}}{2 P_\alpha} X , \hspace{1 cm} X  = \frac{\langle I_- \rangle ^2_{RMS}}{\left(\Delta  I_- \right) ^2 }.
    \label{eq:etaestimate_sup}
\end{equation}
and to propagate the associated uncertainties.

At the beginning of a run we perform a \emph{both-blocked} step. The LO and signal are blocked in order to acquire the ESA \emph{electronic-noise} trace, which captures the additive noise floor of the receiver electronics and the analyzer readout. In the same step, the signal-power monitor (lock detector) is covered and the samples are recorded to determine its dark offset; the mean value of this dark batch is subtracted from all subsequent power-monitor measurements in that run.

After this initialization, we repeat the following acquisition sequence for each dataset:
\begin{enumerate}
    \item \textit{Shot-noise reference (LO only).} The signal path is blocked using the shutter while the LO remains unblocked. After a short settling time, the ESA trace around the IF is acquired with fixed center frequency, span and RBW, and with a fixed number of averages. This trace provides the shot-noise variance (including the electronic noise floor) used in the estimator.
    \item \textit{Quadrature/beat-note measurement (signal + LO).} The shutter is opened to unblock the signal, again allowing a short settling time. A second ESA trace is acquired with identical settings. This spectrum contains the heterodyne tone at the IF together with the noise background and is used to extract the beat-note power.
    \item \textit{Synchronous signal-power measurement.} While the signal remains unblocked, we record samples from the power-monitor detector. The stored voltage samples are corrected by subtracting the previously measured dark offset and are converted to optical power using the calibration described in Sec.~\ref{sec:radiometric_traceability}.
\end{enumerate}
The electronic-noise trace acquired in the initial both-blocked step is used to correct the shot-noise trace in post-processing by subtraction in linear units. Throughout a measurement series we keep the ESA configuration fixed (center frequency at the chosen IF, span, RBW, and averaging count), and we keep the LO power constant; only the signal power is varied. Each acquisition is saved as a single file containing the frequency axis, the electronic-noise, shot-noise, and quadrature traces, the corrected voltage, and the relevant metadata needed for repeatability and uncertainty evaluation.

\section{Traceability and instrument characterization}

\subsection{Radiometric traceability of signal power}\label{sec:radiometric_traceability}

\begin{figure}[ht]
    \centering
    \includegraphics[width = 0.75\textwidth]{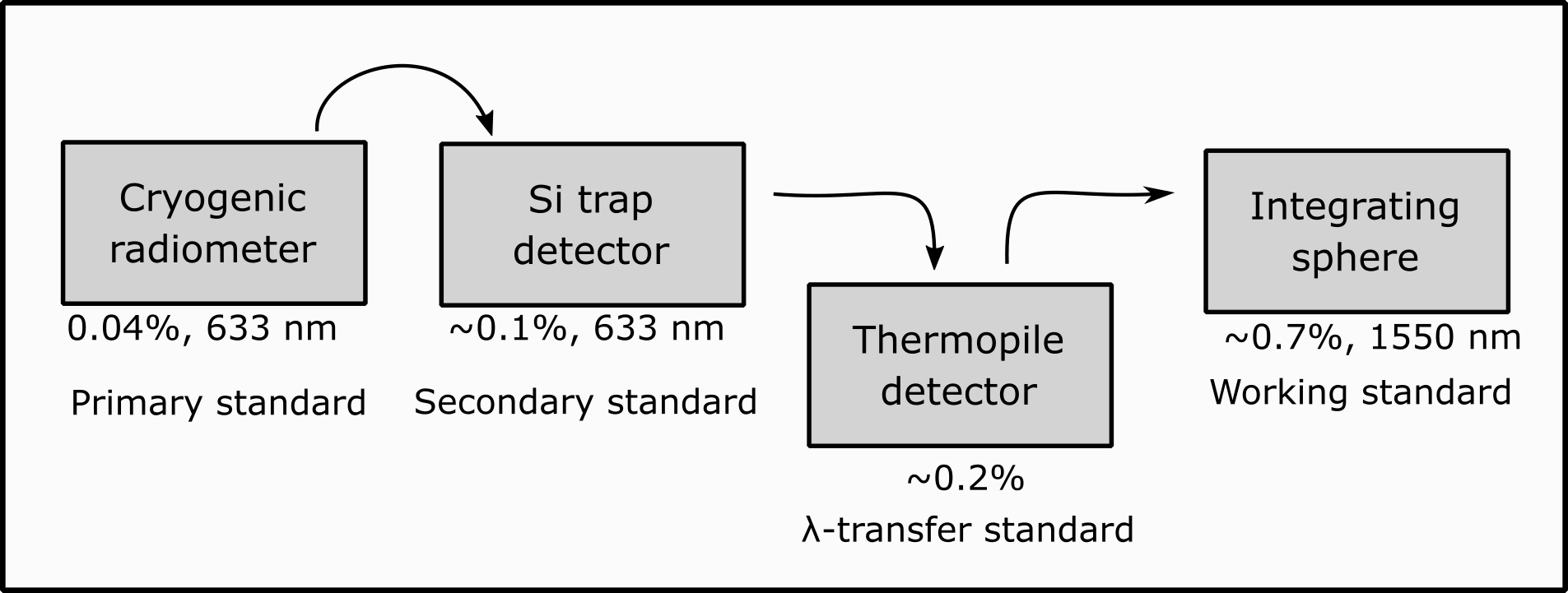}
    \caption{\textbf{Radiometric chain of traceability.} Measurement of optical power is made traceable to the primary radiometric standard, the cryogenic radiometer, by a chain of inter-detector calibrations. Using the flat response (in wavelength) of the metrology-grade thermopile detector, traceability of optical power is transferred from the visible (silicon trap detectors) to the near infrared (integrating spheres with germanium detectors or cooled InGaAs detectors).}
    \label{fig:opticaltraceability}
\end{figure}

The protocol requires an absolute value for the signal optical power $P_\alpha$ (or, equivalently, the photon flux) at the reference plane used for the efficiency estimate. In our implementation, $P_\alpha$ is made traceable through DFM’s radiometric calibration chain to the primary standard, the cryogenic radiometer, which relies on the electrical–optical substitution principle~\cite{martinCryogenicRadiometerAbsolute1985,datlaCharacterizationAbsoluteCryogenic1992}. Figure~\ref{fig:opticaltraceability} summarizes this traceability chain and the role of the different transfer and working standards used to bring the calibration to the near-infrared wavelength region relevant here.

In practice, the signal power entering the heterodyne receiver is too low to be measured directly with a calibrated power meter. Instead, we measure the signal power with a monitor detector and convert the measured voltage to optical power using two calibrated quantities: (i) the attenuation factor $l$ of the fixed ND
filter used to reach the nanowatt regime, and (ii) the responsivity $R$ of the monitor detector in units of \si{\volt\per\micro\watt}. The calibrated signal power is then obtained as
\begin{equation}
P_\alpha \equiv P_{\mathrm{cal}} = V \times \frac{l}{R}\times 10^{-6},
\label{eq:Pcal_traceability}
\end{equation}
where $V$ is the voltage recorded during each acquisition (after subtraction of the dark offset measured at the start of the run). This expression makes it explicit where radiometric traceability enters the protocol: it enters only through $P_\alpha$ (via $l$ and $R$), while the spectral estimator is determined independently from the ESA measurements.

The attenuation factor $l$ is calibrated by measuring optical power before and after the attenuator using a powermeter that is itself calibrated within DFM’s traceability chain. We repeat the before/after measurement several times to obtain a Type-A contribution for $l$, and we include the powermeter calibration uncertainty as a Type-B contribution (appearing twice because both the ``before'' and ``after'' readings are traceable measurements). The monitor responsivity $R$ is calibrated in an analogous way by recording the monitor voltage $V$ while measuring the corresponding optical power with the calibrated powermeter, again combining repeatability with the instrument calibration uncertainty (and a small contribution from the voltmeter). The resulting calibration uncertainties for $l$ and $R$ dominate the uncertainty of $P_\alpha$ in the nanowatt regime, while the repeatability of the voltmeter contributes only weakly.

For completeness, the independent loss-chain estimate used for validation in Sec.~4 is obtained by separately measuring the transmission of the optical path between the power reference plane and the balanced detector, and combining it with the independently measured photodiode quantum efficiencies and the mode-overlap factor.

\subsection{ENBW calibration of the spectrum analyzer}

Throughout this work, the shot-noise reference is obtained from an ESA trace acquired with a finite RBW. To interpret the measured noise level quantitatively, the RBW setting must be converted into the ENBW of the RBW filter. The reason is that the RBW filter is not an ideal rectangular filter: the ESA noise readout corresponds to the noise power transmitted through the actual filter shape, so the effective integration bandwidth differs from the nominal RBW value. In the estimator, this enters through the bandwidth normalization used to convert the measured noise level into a noise power spectral density.

\begin{figure}
    \centering
    \includegraphics[width=0.75\linewidth]{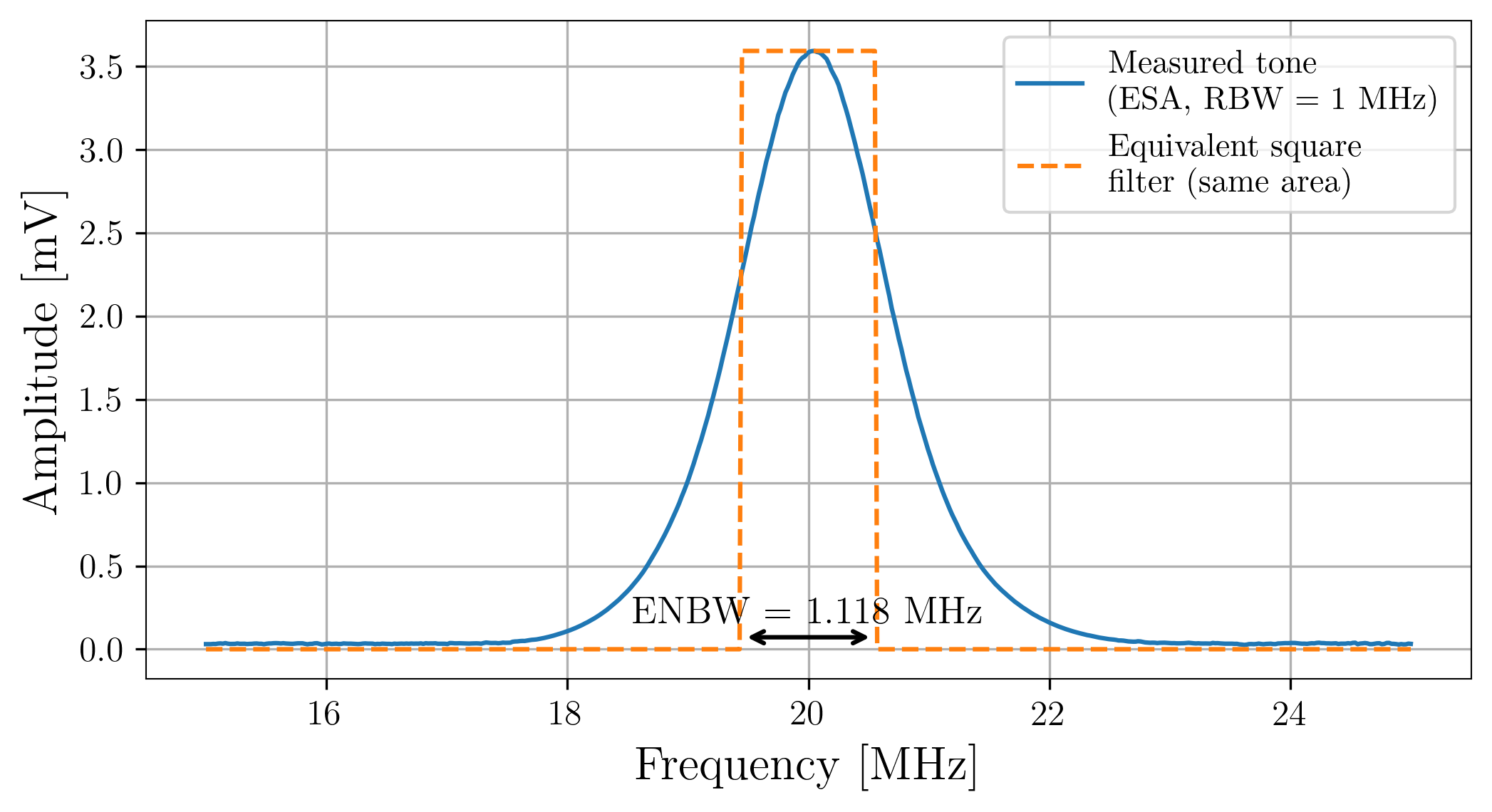}
    \caption{\textbf{Equivalent noise bandwidth (ENBW) calibration of the ESA RBW filter.} Blue curve: measured narrow-tone response of the RBW filter for the RBW = \SI{1}{\mega\hertz} setting (shown in linear voltage units). Orange dashed curve: equivalent rectangular filter with identical integrated squared response. The resulting ENBW = \SI{1.12}{\mega\hertz} corresponds to an ENBW/RBW ratio of 1.12 for the analyzer settings used throughout this work.}
    \label{fig:ENBW}
\end{figure}

We determine the ENBW for the ESA configuration used in the efficiency measurements by measuring the RBW filter response around a narrow tone and constructing an equivalent rectangular filter with the same integrated squared response. Concretely, we set the ESA to the same center frequency, span, and RBW as in the heterodyne measurements (RBW = \SI{1}{\mega\hertz} in the data shown here) and record the trace of a single, spectrally narrow tone. The trace is converted from the ESA’s logarithmic voltage unit (dBmV) to linear voltage (mV). We then normalize the trace to its peak value and compute the ENBW as
\begin{equation}
B_{\mathrm{neq}} \equiv \mathrm{ENBW} = \int \left|\frac{H(f)}{H(f_0)}\right|^2 \mathrm{d}f,
\label{eq:enbw_def}
\end{equation}
where $H(f)$ is the measured RBW filter response and $f_0$ is the tone frequency. Figure~\ref{fig:ENBW} illustrates the measured filter response (blue) and the equivalent rectangular filter (dashed) with width equal to the computed ENBW.

Using this procedure we obtain $\mathrm{ENBW} = \SI{1.12}{\mega\hertz}$ for the RBW = \SI{1}{\mega\hertz} setting used in this work, corresponding to an ENBW/RBW ratio of 1.12. The \SI{0.3}{\percent} uncertainty of the ENBW factor is estimated from repeated acquisitions under unchanged analyzer settings (and from the sensitivity to the fitting/normalization procedure used to locate and normalize the tone).  

\section{Uncertainty analysis} 

\subsection{Main contributors and propagation}

The protocol yields an estimate $\eta$ from Eq.~(\ref{eq:etaestimate_sup}), based on three experimentally determined quantities: the dimensionless ratio $X$ extracted from the ESA spectra, the calibrated signal power (or photon flux) $P_\alpha$, and the analyzer equivalent noise bandwidth $B_{\mathrm{neq}}$. The combined standard uncertainty is propagated in accordance with \cite{GuideExpressionUncertainty2023} using Eq.~(5). In practice, it is convenient to express it in relative form:

\begin{equation}
\frac{u^2(\eta)}{\eta^{\,2}}
=
\frac{u^2(X)}{X^{2}}
+
\frac{u^2(P_\alpha)}{P_\alpha^{2}}
+
\frac{u^2(B_{\mathrm{neq}})}{B_{\mathrm{neq}}^{2}},
\label{eq:eta_unc_rel}
\end{equation}
where each term corresponds to an experimentally identifiable contribution:

\begin{itemize}
    \item \textit{Uncertainty in calibrated signal power, $u(P_\alpha)$.} The signal power is obtained from the monitor-detector voltage measurement and the calibrated attenuation and responsivity factors. In our implementation,
    \begin{equation}
        P_{\mathrm{cal}} = V \times \frac{l}{R}\times 10^{-6},
    \end{equation}
    so that $u(P_\alpha)$ is dominated by the calibration uncertainties of the fixed ND filter attenuation $l$ and the detector responsivity $R$, with a smaller contribution from the repeatability of the voltage readout $V$.

    \item \textit{Uncertainty in the ENBW, $u(B_{\mathrm{neq}})$.} The bandwidth normalization enters through $B_{\mathrm{neq}}$, which corrects the ESA resolution bandwidth (RBW) to a equivalent noise bandwidth. The ENBW determination is based on the measured filter response (Supplement~1), and the associated uncertainty is treated as a Type~B contribution in the present work.

    \item \textit{Repeatability of the spectral estimator, $u(X)$.} The quantity $X$ is extracted from the ESA spectra by evaluating the beat-note and noise levels in a fixed frequency window around the tone. The corresponding uncertainty is obtained as a Type~A contribution from repeated acquisitions under identical conditions, and it captures short-term fluctuations of the optical alignment, laser offset lock, and electronic readout.

\end{itemize}

\subsection{Dominant uncertainty regimes and what limits performance}

In the free-space measurements at \SI{20}{\mega\hertz}, the uncertainty budget is typically dominated by $u(P_\alpha)$, i.e. by the calibration of the fixed attenuation $l$ and responsivity $R$ used to convert the monitor voltage to optical power. The ENBW term $u(B_{\mathrm{neq}})$ is smaller but non-negligible and provides a constant systematic contribution common to all points once the ESA settings are fixed, while the repeatability contribution $u(X)$ is generally minor for the averaging settings used here. This hierarchy is consistent with the fact that the spectrum-based estimator is comparatively stable once the LO power and IF settings are fixed, whereas the absolute efficiency value is set by how well the signal photon flux is known.

For the attenuation-validation measurements using calibrated ND filters, an additional uncertainty enters through the transmission factor $T$ used to scale the independent loss-chain prediction. In some cases the expanded uncertainty of $T$ becomes comparable to, or larger than, the uncertainty associated with $P_\alpha$ in the protocol measurement. This is most pronounced for the higher-transmission filters, where the transmission calibration is limited by repeatability and by the power-meter calibration contributions from the ``before'' and ``after'' measurements. As a result, the uncertainty band of the independent prediction $\eta_{\mathrm{sep}}(T)$ broadens for selected attenuation points even though the protocol-derived $\eta$ retains a similar uncertainty. 

%Manual citation